\documentclass[11pt, oneside]{elsarticle} 
\usepackage{geometry}                		
\usepackage{graphicx}				
\usepackage{amssymb}

\begin{document}

\title{Photometric Limits on the High Resolution Imaging of Exoplanets Using the Solar Gravity Lens}
\author{Phil A. Willems}
\ead{pwillems@jpl.nasa.gov}
\address{Jet Propulsion Laboratory, California Institute of Technology, 4800 Oak Grove Drive, Pasadena, California, USA 91109}

\begin{abstract}
We present an analysis of high-resolution imaging of an exoplanet by a meter-class telescope positioned at a real image of the exoplanet created by the solar gravity lens. We assume an exoplanet viewed in full phase and a simple deconvolution method to correct for the intrinsic blur caused by aberrations in the solar gravity lens, and account for the foreground light of the solar corona. We derive equations for the measurement times required for the telescope to produce such a high-resolution image, and find that due to shot noise considerations these times are large compared to human lifespans. We also discuss how measurement times could potentially be significantly shorter for exoplanets with special orbital and rotational parameters. 
\end{abstract}


\maketitle

\section{Introduction}

Gravitational lensing of light is a well-known phenomenon that has been widely used to gather images of distant galaxies and measure the masses of the intervening lensing galaxy clusters.~\cite{tr:ARAA} Gravitational lensing has also been used to detect and characterize exoplanets around distant stars, by the study of the transient amplification of light seen at Earth from background stars when the exoplanet and its host star passes directly in front of them.~\cite{ga:Exoplanets} It is perhaps less well-known that the gravitational field of a star like the Sun produces a real image of distant exoplanets that could potentially be measured. These real images produced by the Sun are to be found in the distant reaches of our solar system, several hundred astronomical units (au) and farther away. The prospect of collecting a high-resolution image of an exoplanet by using the solar gravity lens (SGL) as the primary aperture of an astronomically large telescope is enticing. Some published reports suggest that images with megapixel spatial resolution can be collected in this way.~\cite{tu:NIAC, tu:SciAm}

Recently, G. Landis pointed out some of the challenges to making a resolved image of an exoplanet using the SGL, in particular the foreground light of the solar corona and the blurriness inherent in the SGL itself.~\cite{la:AIAA} This article extends upon Landis's work by exploring the photometric limits to high-resolution imaging of exoplanets by use of the SGL. In particular, it considers the signal-to-noise ratios necessary to to produce an image of a given resolution and fidelity, given the spherical aberration in the SGL and the noise signal introduced by the solar corona. 

\section{Imaging of Distant Objects Using the Solar Gravity Lens}

There exists an extensive literature that describes the optics of a gravitational lens, which we do not attempt to reproduce here, presenting only those results that are most relevant to imaging and photometry.

The mass of the Sun curves the path of light passing it, effectively making it a lens, albeit one with significant spherical aberration. Consider a point source of light behind the Sun and distant enough that it illuminates the Sun with essentially a plane wave, and a viewer at an image plane in front of the Sun at a distance \( z \). That viewer will see the light focused to its location from an Einstein ring surrounding the Sun with a radius \( r _E \) given by
\begin{equation} \label{eq:Einsteinring}
r_E = \sqrt { \frac{4GM_{\odot }z} {c^2}} 
\end{equation}
where \( G \) is the gravitational constant, \( M _{\odot } \) the solar mass, and \( c \) the speed of light.~\cite{la:AIAA} Because the Einstein ring must be wider than the Sun to not be blocked by the Sun's photosphere, \( z > 550 \) au. Each point on the surface of an exoplanet will be brought to a focus somewhere in the image plane, producing a real image there. By simple geometry, the image will be demagnified by \( d/z \), where \( d \) is the distance from the Sun to the exoplanet. For an Earth-sized exoplanet located between one and ten parsecs away from the Sun and observed at about 1,000 au from the Sun, the image is kilometers in size and cannot be measured all at once by a meter-scale telescope.

\subsection{Estimating the Blurriness of a SGL Exoplanet Image} \label{sec:blur}

The spherical aberration inherent in the SGL appears as the variation of Einstein ring radius with focal distance in Equation \ref{eq:Einsteinring}. Its severity may be judged by noting that the image plane distance \( z \) grows without limit as the Einstein ring radius increases, that is, for larger impact parameters at the Sun. For any chosen image plane beyond 550 au, the focused image produced by the 'correct' Einstein ring is obscured by all the defocused images from the 'incorrect' Einstein rings. 

Turyshev and Toth~\cite{tu:PRD} define the gain of the SGL to be the increase in intensity of the light at the image plane over that incident in a plane wave at the Sun. This gain $\mu _z$ is very strongly dependent on transverse displacement from the line along the plane wave direction through the center of the Sun, and is given by
\begin{equation} \label{eq:gain}
\mu _z = \frac{4 \pi ^2}{1 - e^{-4 \pi ^2 r_g/\lambda}} \frac{r_g}{\lambda}J ^2 _0 \left( 2 \pi \frac{\rho}{\lambda} \sqrt{\frac{2r_g}{z}}\right)
\end{equation}
where \( \rho \) is the transverse displacement, \( r_g =  2GM_{\odot }/c^2 \) is the Schwarzschild radius of the Sun, and \( \lambda \) is the light wavelength. Throughout this article, we will assume that \( \lambda =1 \mu\)m for convenience. The on-axis gain \( ( \rho = 0) \) is independent of \( z \) and for \( \lambda =1\) \(\mu\)m is \(1.17 \times 10^{11}\). This beam pattern is surprisingly highly focused- the first minimum in the SGL beam pattern occurs only about 5 cm from the central maximum- but on average the gain decreases no faster than inversely with the transverse displacement beyond the first few minima. This is a much slower rolloff than seen for the Airy disk point spread function of an unaberrated circular lens.

This relatively slow drop in gain with transverse displacement is the source of the blur in the image. The exoplanet light reaching the telescope is not a single plane wave, but a superposition of plane waves from the entire face of the exoplanet, and so the telescope actually observes a superposition of Einstein rings. These are not resolvable from one another by a meter-class telescope, which sees all the rings as essentially overlapping. Thus, light from all points on the exoplanet surface contribute significantly to the intensity seen at any one point in the image.

To estimate the magnitude of the blur, we determine what fraction of the light incident at the central region of the image comes from light outside that region on the exoplanet.

 Assume that the exoplanet is a uniform disk, so that its surface brightness $B_s = B_0$ for $r \le r_2$, $B_s = 0$ for $r > r_2$ where $r$, $\theta $ are polar coordinates at the exoplanet and $r_2$ is the exoplanet radius. Let $\rho $, $\phi $ be the corresponding coordinates in image plane at the SGL. We collect the light at the center of the image with a telescope that has aperture of radius $r_1 \ll r_2$. The power $P_1$ collected by the telescope from within $r_1$ on the exoplanet is
\begin{equation} \label{eq:P1}
P_1 \propto \int _0^{2\pi } d\phi \int _0^{r_1} \rho d\rho \int _0^{2\pi } d\theta \int _0^{r_1} rdr J_0^2(k|\vec r - \vec \rho|) B_s(r)
\end{equation}
and the power $P_2$ collected from outside $r_1$ is
\begin{equation} \label{eq:P2}
P_2 \propto \int _0^{2\pi } d\phi \int _0^{r_1} \rho d\rho \int _0^{2\pi } d\theta \int _{r_1}^{r_2} rdr J_0^2(k|\vec r - \vec \rho |) B_s(r)
\end{equation}
with the same proportionality, where  $k = 2\pi \sqrt{2 r_g /\lambda ^2 z}$. In Eqs.~\ref{eq:P1}-~\ref{eq:P2} we have suppressed the unimportant image magnification factor. An approximate solution to Eq.~\ref{eq:P1} can be found by replacing the uniform exoplanet surface brightness for $r \le r_1$ with a delta function at $r=0$ having the same total power $\pi r_1^2 B_0$. This yields the result
\begin{equation} \label{eq:P1approx}
P_1 \propto \pi r_1^2 B_s \int _0^{2\pi } d\phi \int _0^{r_1} \rho d\rho J_0^2(k\rho ).
\end{equation}
We can likewise find an approximate solution to Eq.~\ref{eq:P2} by noting that over most of the region $r_1 < r \le r_2$, $r \gg r_1 > \rho$, and so $k|\vec r -\vec \rho | \approx kr$, giving the estimate
\begin{equation} \label{eq:P2approx}
P_2 \propto \pi r_1^2 B_s \int _0^{2\pi } d\theta \int _{r_1}^{r_2} rdr J_0^2(kr). 
\end{equation}
The integrals in Eqs.~\ref{eq:P1approx}-\ref{eq:P2approx} are easily solved, yielding
\begin{equation}
{ {P_2} \over {P_1} } \approx { {\pi r_2^2 \left[ J_0^2 \left( k r_2 \right) + J_1^2 \left( k r_2 \right) \right]
  -  \pi r_1^2 \left[ J_0^2 \left(k r_1 \right) + J_1^2 \left(k r_1 \right) \right]} \over {\pi r_1^2 \left[ J_0^2 \left( k r_1 \right) + J_1^2 \left( k r_1 \right) \right]} } \approx { {r_2} \over {r_1} },
\end{equation}
where the latter equation holds because, asymptotically as \( x \rightarrow \infty \), the Bessel functions \(J_n(x)\) are sinusoids with amplitude falling as \( 1/\sqrt{x} \).  The blur contribution to the power within $r_1$ is thus \(\sim r_2/r_1 \) times larger than focused light there. Away from the center of the image, the blur contribution is less than this, and is about 60\% as large at the very edge of the image, as we will show in a later section.

\subsection{The Intensities of the Einstein Ring and Solar Corona as Seen at the SGL}

We cannot simply put a photodetector at the image plane of the SGL and collect the exoplanet light falling upon it. The light from the Sun at that location will be many orders of magnitude brighter. We therefore assume the imager to be a telescope with a coronagraph that blocks the sunlight to an ignorably small level while allowing light from the Einstein ring to reach the telescope without appreciable attenuation. We do not consider here the technical feasibility of such a coronagraph. We also assume that the telescope's aperture has diameter \( D_{tel} = 1\) m. This is well within the size range of current optical space telescopes. This choice is arbitrary, but the results derived in this article can easily be reanalyzed for other sizes of telescope aperture. We let the telescope detector have perfect quantum efficiency, and no read or dark noise.

\begin{figure}
  \includegraphics[width=\linewidth]{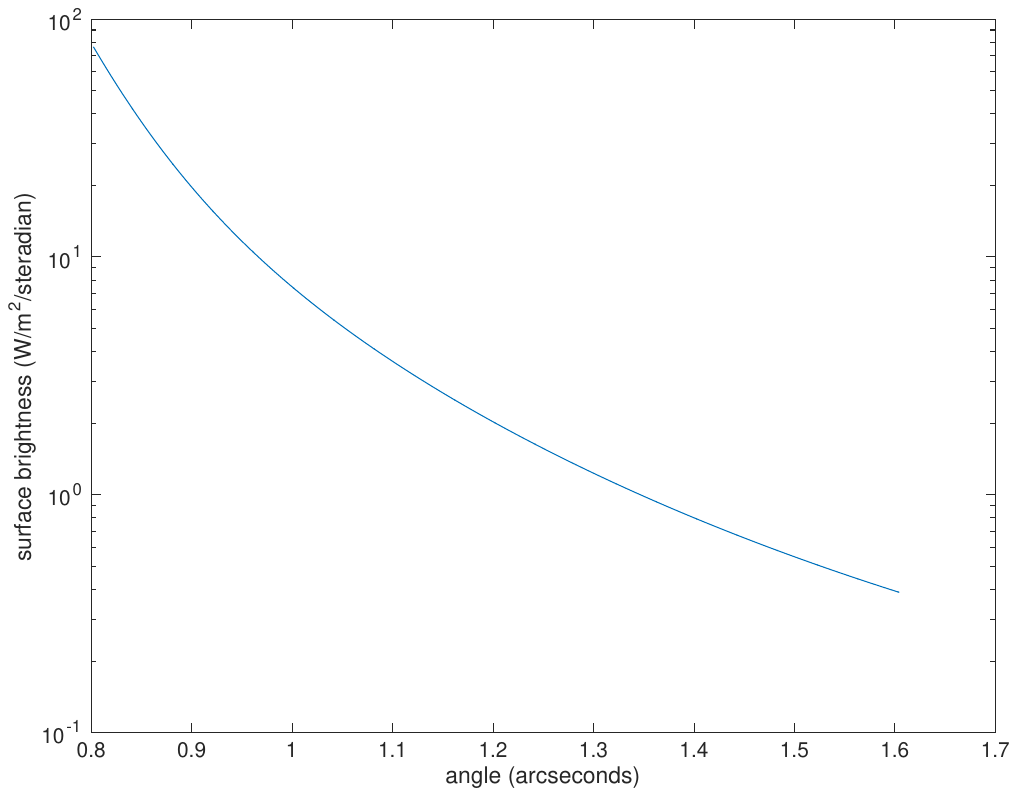}
  \caption{The typical surface brightness of the solar corona as seen at 1200 au.}
  \label{fig:coronabrightnessatSGL}
\end{figure}

The occulter cannot also suppress the light from the Sun's corona, because the Einstein ring shines through the corona on its way to the image plane. The solar corona's brightness and spatial features are constantly changing, and its appearance from the image plane will likely differ from its appearance at Earth due to the extended distribution of the F-corona and the possible location of the image well outside the ecliptic. However, we can take this representative form~\cite{go:TheSolarCorona} for its surface brightness:
\begin{equation}
B_{cor}(r) \sim 18.9 \times \left( \frac{0.0532}{r^{2.5}} + \frac{1.425}{r^7} + \frac{2.565}{r^{17}} \right) { {\rm W} \over {{\rm m}^2 \cdot {\rm steradian}} },
\end{equation}
where \( r\) is in units of solar radii. This surface brightness distribution strictly applies only to the K-corona, which dominates the brightness within 2 solar radii. Figure~\ref{fig:coronabrightnessatSGL} shows the radial brightness distribution of the corona as seen at a distance of \(z =1200\) au, a distance typical of proposed SGL missions. For this distribution, the coronal light power collected by the telescope is \(P_{cor} = 6.78 \times 10^{-10}\) W, with a photon rate of \(Q_{cor} = 3.41 \times 10^9\) photons per second.

We make some specific assumptions about the exoplanet itself. First, we assume it to be an Earth analog- it has the same radius \(R_{exo} = 6370\) km, average albedo \(a = 0.3\), and instellation \( I=1\) \({\rm kW} / {\rm m}^2 \) as Earth. We also assume it to be fully in phase. We allow its distance $d$ from the Sun to vary, but require that \(d \ge 1.3\) parsecs, which is the distance to the nearest star, Proxima Centauri.

We can now estimate the detected optical power from the Einstein ring. Let the {\it directly imaged} region be that part of the exoplanet from which a line projected through the center of the Sun falls within the telescope aperture. The maximum of the SGL optical gain pattern will fall within the telescope aperture for light originating within this directly imaged region, and this light is the desired image signal.  By simple geometry, this region has projected radius \( r_{dir} = D_{tel}d/2z\). The power radiated toward the Sun by the directly imaged region is $aI\pi r_{dir}^2/2 \pi d^2$, where the radiated intensity falls off with distance as \( 1/2\pi d^2 \) rather than \( 1/4\pi d^2 \) because the illuminated portion of the exoplanet surface radiates mostly into the hemisphere of its daytime sky. Turyshev and Toth have derived the average intensity gain for a beam centered on the telescope aperture to be~\cite{tu:PRD}

\begin{equation}
{\overline \mu }_z = { {4\pi ^2} \over {1-e^{-4\pi ^2 r_g/\lambda} }}   {{r_g} \over {\lambda}}    \left\{ J_0^2 \left( \pi { {D_{tel}}\over {\lambda} } \sqrt{ 2r_g \over z}  \right)  +  J_1^2 \left( \pi { {D_{tel}}\over {\lambda} } \sqrt{ 2r_g \over z}  \right) \right\}.
\end{equation}

At $z=1200$ au, ${\overline \mu }_z = 4.14 \times 10^9$. We make the simplifying assumption that this average gain applies everywhere in the directly imaged region, and not just at its center. Thus, the directly imaged power $P_{dir}$ is

\begin{equation}
P_{dir} = { {\pi D_{tel}^2 {\overline \mu }_z} \over 4 } { {aI\pi r_{dir}^2} \over {2 \pi d^2} }.
\end{equation}

The rest of the exoplanet disk outside the directly imaged region also contributes to the detected exoplanet optical power, through the blur. For \( D_{tel} = 1\) m, \( r_{dir}\) is 133 m for an exoplanet orbiting Proxima Centauri. This is much less than the 6370 km radius of the exoplanet. Even for exoplanets ten or a hundred times more distant, only a small fraction of the exoplanet is directly imaged, so the estimate that the blur power at the center is \(r_2/r_1\) times greater than the directly imaged power remains valid. The total exoplanet optical power $P_{exo}$ is therefore $R_{exo}/r_{dir}$ times the directly imaged power $P_{dir}$, or

\begin{equation}
P_{exo} = { {\pi D^2_{tel} \overline \mu _z} \over {4} }  \frac{aI\pi r_{dir}^2}{2\pi d^2}\frac{R_{exo}}{r_{dir}} = { {\pi \overline \mu _z aI D^3_{tel} R_{exo}} \over {16 d z} } 
\end{equation}

Here we have made the slightly conservative assumption that the amount of blur everywhere in the image is like that at its center, which offsets the   slightly nonconservative assumption that the gain everywhere in the directly imaged region is the same as at its center. Note that the total exoplanet light power collected by the telescope is inversely proportional to the exoplanet distance \(d\). There is a subtle reason for this. The intensity of the light radiated by the exoplanet falls off as \(1/d^2\). But the area directly observed by the telescope increases as \(d^2\). These two effects therefore cancel, as is typical for resolved objects. Whence the \(1/d\) dependence then? The area of the exoplanet outside the directly imaged region also drops with greater exoplanet distance- there is less 'extra exoplanet' to blur into the pixel of interest. For distances large enough that the telescope directly images the entire exoplanet, there is no blur contribution, the directly imaged region can no longer increase with distance, and so the power received by the telescope then falls off as \(1/d^2\). For exoplanet distances close enough that the exoplanet can be resolved by the 1 m telescope at the SGL,
\begin{equation}
P_{exo} = 2.11 \times 10^{-13} {\rm W} \left( {1.3 {\rm pc} \over d} \right)
\end{equation}
with a photon rate \( Q_{exo} = 1.06 \times 10^{6}/d \) per second.

\begin{figure}
  \includegraphics[width=\linewidth]{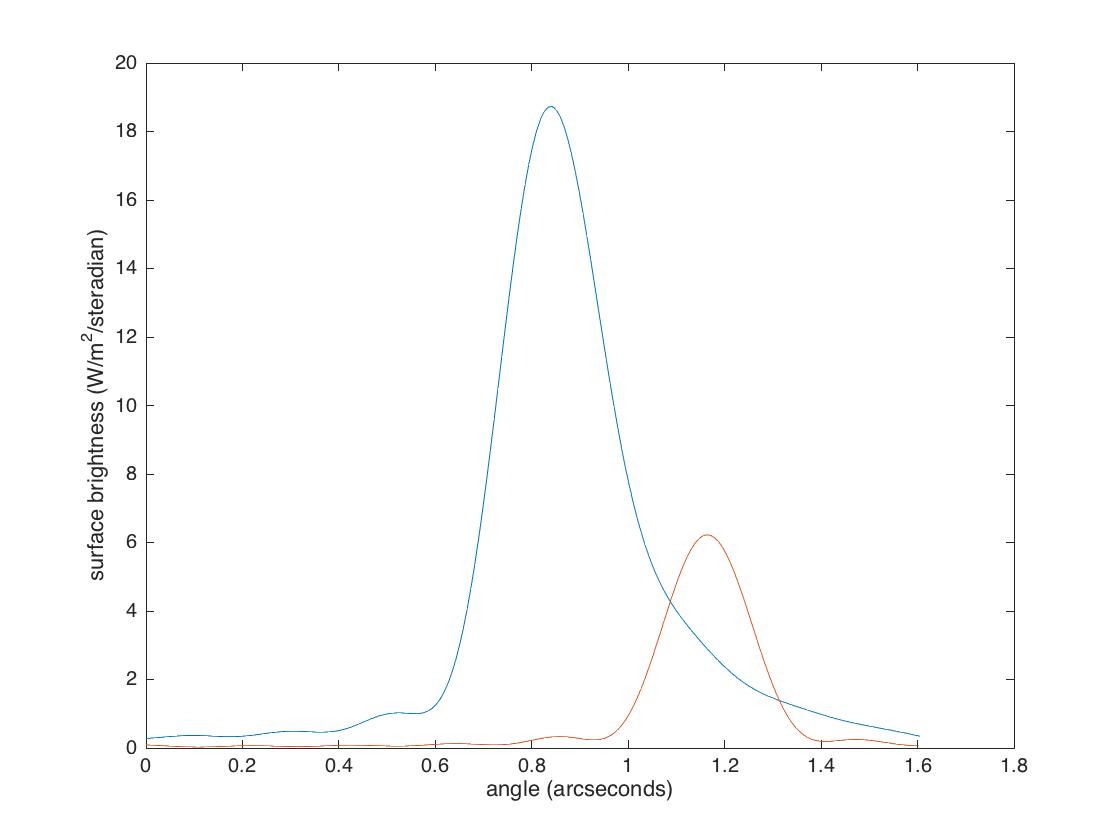}
  \caption{The surface brightnesses of the solar corona (blue) and the exoplanet (red), convolved by the spatial resolution of a one-meter diameter telescope at 1,200 a.u. The occulter is assumed not to affect the telescope point spread function outside its dark zone. The exoplanet's Einstein ring surface brightness is magnified 1000x relative to the corona.}
  \label{fig:coronavsexoplanet}
\end{figure}

It remains to determine to what extent the exoplanet light in the Einstein ring can be separated from the coronal foreground light. The Einstein ring surrounds the Sun and so its radius is resolved by a 1-m telescope at the SGL. The Einstein ring's width is comparable to the angular extent of the exoplanet as seen from the Sun, and so will not be resolved. We have convolved the Einstein ring with the point spread function of a round 1m aperture and plotted it along with the corona brightness distribution in Figure \ref{fig:coronavsexoplanet}. The radial location of the Einstein ring is that for \( z = 1200 \) a.u.; for larger \( z \) the exoplanet peak shifts to the right. Note that in the figure the exoplanet signal is multiplied 100x in order to make it visible.

Clearly, the corona overlaps substantially with the Einstein ring at this SGL location. The fraction of the coronal light power falling between the first minima in diffraction-limited Einstein ring signal is about 35\%. This fraction of the coronal light cannot be resolved from the Einstein ring in the telescope image, and must be subtracted off photometrically.

\subsection{Deconvolving the Blurry Measurements to Form a High-Resolution Image}

Creating a megapixel image requires at least one million separate measurements. For a typical photograph, each detector pixel within the camera is performing a separate measurement. This is not the case for exoplanet imaging at the SGL. Only the pixels in the telescope detector that image the Einstein ring are measuring the exoplanet, and the observed Einstein ring contains a blend of light from the entire exoplanet surface, due to the blur of the SGL, along with a large coronal foreground signal. Substantial deconvolution would be required to synthesize the exoplanet image.

We adopt here a simple scenario, though admittedly an unphysical one, in order to enable a straightforward measurement and deconvolution method that we can then use to estimate required signal to noise ratios and, finally, integration times at the SGL. Namely, we assume that the planet is always in phase, that it is not rotating, and that its features never vary with time (e.g. no clouds passing, no seasons). No real exoplanet would provide such an unchanging aspect, but other potential extended sources of interest like nebulae and galaxies would. We also assume that the telescope can track the exoplanet's orbit and maneuver around the exoplanet image at the SGL at will. Finally, we assume that the image of the solar corona can be predicted by independent means- that is, without using additional SGL telescope time, perhaps using another nearby telescope outside the SGL image- and subtracted from the Einstein ring image with shot noise limited precision. 

Given these assumptions, we consider a simple method to form a megapixel image of the exoplanet:
\begin{enumerate}
\item Divide the SGL image plane into a 1000x1000 grid that just contains the exoplanet in the 'directly imaged' sense.
\item Position the telescope in the center of each pixel for a time $T_1$ and measure the total combined power of the Einstein ring and solar corona.
\item Subtract the independently determined coronal power for each pixel to get the Einstein ring contribution at that pixel.
\item Multiply the vector of image pixels so obtained by a deconvolution matrix describing the SGL blur to obtain a vector of object pixels at the exoplanet.
\end{enumerate}

This concept can be understood mathematically as follows. We represent the source object by a vector $\vec S$ of one million elements, one for each pixel, and the resulting image at the SGL by a similar vector $\vec R$, where the source pixel $S_i$ is directly imaged by $R_i$ for all $i$. The blur is represented by a square convolution matrix $\bf B$, which has one million million elements; each element ${\bf B}_{ij}$ of $\bf B$ contains the gain at the image pixel $i$ from the source pixel $j$. 
If we denote the light from the corona at the image plane by the vector $\vec C$, then the vector of pixel measurements $\vec F$ is given by
\begin{equation}
\vec F = \vec R + \vec C = {\bf B} \cdot \vec S + \vec C.
\end{equation} 
The deblurred exoplanet image $\vec S$ can thus be calculated from the measurements by
\begin{equation}
\vec S = {\bf B}^{-1} \cdot \vec R = {\bf B}^{-1} \cdot (\vec F - \vec C ).
\end{equation}

We assume for the sake of argument that we wish to determine the elements of $\vec S$ with a SNR of 10. We will then need to determine $\vec R$ with a higher SNR; how much higher can be found by inspection of ${\bf B}^{-1}$.

We generated convolution matrices ${\bf B}$ of smaller size than $10^6$x$10^6$ out of practicality. The value of each ${\bf B}_{ij}$ element was determined by numerically integrating the gain (Eq.~\ref{eq:gain}) over a circular telescope aperture positioned at the center of image pixel $i$ for all points within the square source pixel $j$. In our scenario, the telescope aperture is smaller than a single image pixel. Thus, a ${\bf B}_{ii}$ element accounts for the 'intrapixel blur' due to all light that comes from within the 'directly imaged' pixel, including that not directly imaged by the smaller telescope aperture. For concreteness, we chose an image pixel size of 12.6 m for all sizes of ${\bf B}$ we studied. An exo-Earth at 5 parsecs distance observed using the SGL 1,000 a.u. from the Sun would just fill such a megapixel image. The smaller than $10^6$x$10^6$ matrices we investigated thus corresponded to smaller than Earth-sized exoplanets at the same distance, imaged with the same pixel size at the source, or exo-Earths at greater distance, with correspondingly larger pixels at the source. For our calculations, we used a raster-scanned pixel ordering for $\vec S$, $\vec R$ and $\vec F$ out of convenience, but other choices are possible and would not affect the results.

\begin{figure}
  \includegraphics[width=\linewidth]{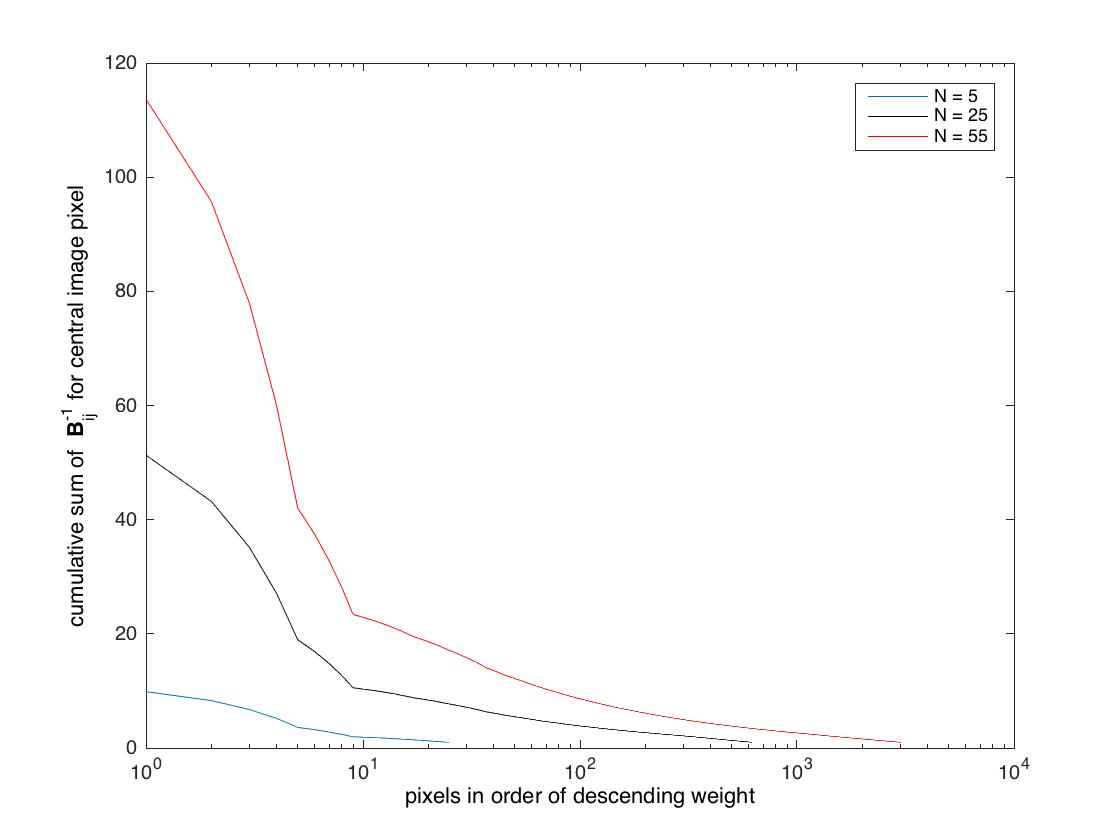}
  \caption{The cumulative sum of the elements in the row of ${\bf B}^{-1}$ corresponding to the central image pixel, for three values of $N$. The matrix elements have been reordered by decreasing relative weight in the sum for clarity, and the sums for all three values of $N$ have been scaled to unity for ease of comparison.}
  \label{fig:Binverseplot}
\end{figure}

\begin{figure}
  \includegraphics[width=\linewidth]{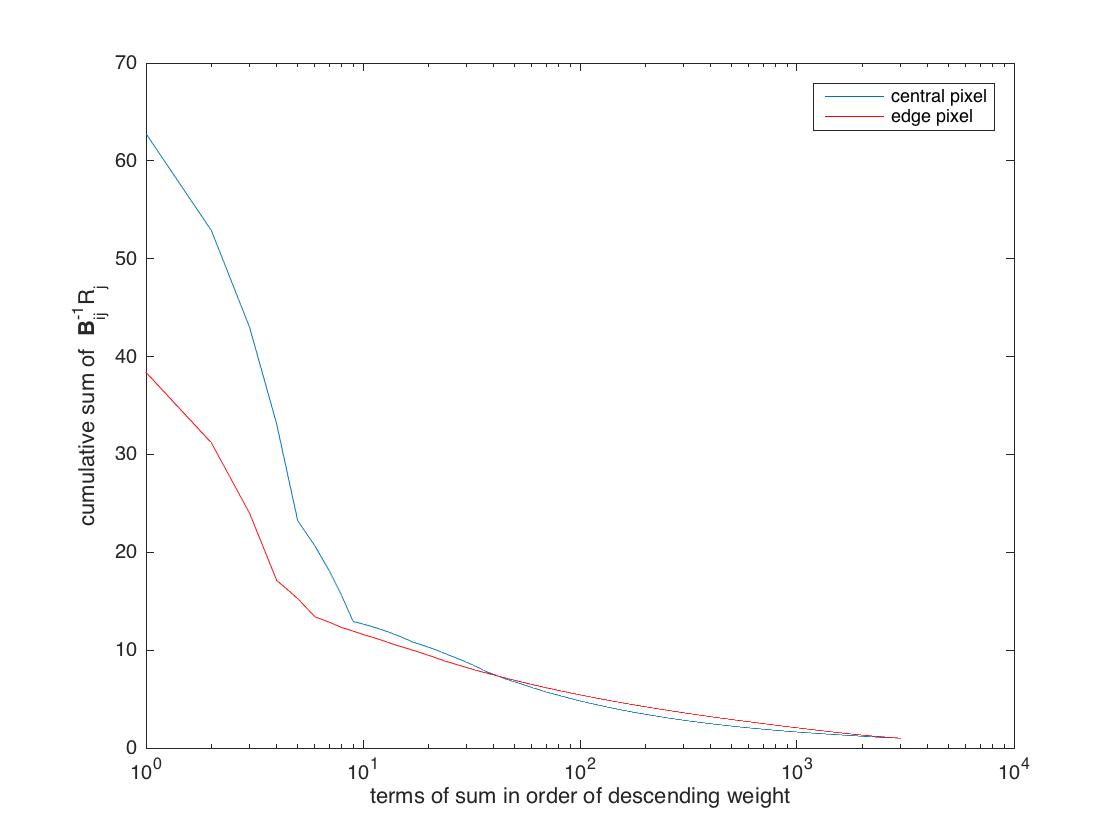}
  \caption{The cumulative sum of the terms in $\sum_{j}{\bf B}_{ij}^{-1}R_j$ corresponding to pixels at the exoplanet center and edge. As in Figure \ref{fig:Binverseplot}, the terms in the sum have been reordered by decreasing relative weight in the sum for clarity. For this calculation, $N=55$.}
  \label{fig:centervsedge}
\end{figure}

Inspection of ${\bf B}^{-1}$ for various sizes shows that it has no null elements: measurements from all pixels must be combined to recover the source brightness at any one pixel. Only the diagonal elements of ${\bf B}^{-1}$ are positive, all other elements being negative. To illustrate the relative scales of the elements of ${\bf B}^{-1}$, Figure \ref{fig:Binverseplot} shows how the sum of all the elements in the row of ${\bf B}^{-1}$ contributing to the central pixel of $\vec S$ is about $1/2N$ smaller than the diagonal element of that row alone, where the exoplanet image is NxN. In making this plot, we have reordered the matrix elements from largest size to smallest. Figure \ref{fig:Binverseplot} reveals that the eight largest off-diagonal elements- i.e., those representing pixels immediately surrounding the central pixel- collectively have a weight about 80\% that of the central pixel. When calculating the SNR of the deconvolved image $\vec S$, what matters is the sum $S_i=\sum_{j}{\bf B}_{ij}^{-1}R_j$. Figure \ref{fig:centervsedge} shows how the measured value at a given pixel is reduced by the measurements at all the other pixels by the deconvolution for pixels at the center and edge of the exoplanet image, for the case that $N=55$. At the image center, the combination all the off-diagonal terms tends to cancel the single diagonal term by a factor of about $1/N$. This behavior was seen for all values of $N$ we studied. This is consistent with the calculation in Section~\ref{sec:blur} which showed that the blur contribution to the central radius $r_1$ region of an SGL image of overall size $r_2$ is $r_2/r_1$  times larger than the directly imaged contribution. The relative contribution of the negative off-diagonal elements falls off for rows of ${\bf B}^{-1}$ corresponding to pixels of increasing distance from the center, to about 60\% less at the edge of the exoplanet disk. 

Taking the blur contribution at the center as representative of the whole image for simplicity, we can therefore say that to get a given ${\rm SNR}_S$ per pixel in the recovered vector $\vec S$ for an $N \times N$ image, we need to know the pixels of $\vec R$ with an ${\rm SNR}_R$ of at least
\begin{equation}
{\rm SNR}_R \ge {\rm SNR}_S \times N
\end{equation}
and we must do this for all $N^2$ pixels in $\vec R$. We do not measure $\vec R$ directly, but get it by subtracting an estimate of the coronal signal $\vec C$ from the actual measurements $\vec F$. We have already established that the corona outshines the Einstein ring by of order a thousand or more at the SGL. Therefore the shot noise of the coronal light dominates in the estimate of ${\rm SNR}_R$. Since $fQ_{cor} \gg Q_{exo}$, we have
\begin{equation}
{\rm SNR}_R = {{Q_{exo}T_1} \over {\sqrt{{fQ_{cor}T_1}}}} = {{ Q_{exo} } \over { \sqrt{ fQ_{cor} } } } \sqrt{ T_1 },
\end{equation}
where $T_1$ is the time of measurement of one pixel. The time required to collect the data for one pixel is then
\begin{equation}
T_1 \ge { {N^2 fQ_{cor} {\rm SNR}^2_S} \over {Q^2_{exo} }  }
\end{equation}
and the time $T_{tot}$ to collect the whole image is
\begin{equation} \label{eq:Ttotal}
T_{tot} \ge { {N^4 fQ_{cor} {\rm SNR}^2_S} \over {Q^2_{exo} }  }
\end{equation}

The time required to produce a 1000x1000 pixel image of an exo-Earth at the distance of Proxima Centauri with an SNR of 10 is by this formula is \( T_{tot} = 3.7 \times 10^{10} \) seconds, or 1,200 years. Because \( Q_{exo} \) is inversely proportional to the exoplanet distance \( d \), an exoplanet twice as far away will take four times as long to image. The integration time for a $N$x$N$ image scales strongly as $N^4$, so reducing the resolution to 500x500 reduces the measurement time by a factor of 16 to 75 years.

We can also calculate the time to collect an exoplanet image if there were no light from the solar corona. In that case, ${\rm SNR}_R = \sqrt{Q_{exo} T_1}$, which leads to the total integration time $T_{tot} \ge N^4 {\rm SNR}^2_S/Q_{exo}$, which is $fQ_{cor}/Q_{exo} \simeq 1000$ times shorter than the case when the coronal light is correctly included, or about three years for a megapixel image of an exo-Earth orbiting Proxima Centauri. It is clear from this that the coronal foreground light dramatically impairs the feasibility of using the SGL for imaging of faint objects.

\section{Effects of the Exoplanet's Orbital and Rotational Motion}

The foregoing discussion clearly implies that megapixel imaging of a static exoplanet is not feasible with a meter-scale telescope within a human lifespan. This analysis can readily be generalized to imaging of other astrophysical sources such a distant galaxies that, unlike exoplanets, truly are static over this measurement time. But an exoplanet is clearly not a static object. It orbits its host star, thus moving and changing its phase. It rotates, thus changing the features it brings into view at the SGL. How might these change the results above?

If one could control the illumination of the exoplanet, a simple method to remove the blur is available. One could illuminate only one pixel on the exoplanet, gather the light for that pixel through the solar corona, and then illuminate the next pixel, and so on, until the whole exoplanet has been imaged. Because all regions contained in pixels other than the one being measured are dark, they cannot contribute light to the pixel of interest, so the blur need not be deconvolved away. In this case \( \vec R = \vec S \), so \( {\rm SNR}_R = {\rm SNR}_S \), and Equation \ref{eq:Ttotal} becomes
\begin{equation}
T_{tot} \ge { {N^2 fQ_{cor} {\rm SNR}^2_S} \over {Q^2_{exo} }  }.
\end{equation}
The time to collect a megapixel image of an exo-Earth at Proxima Centauri would then drop from 1,200 years to only 10 hours. 

Of course, we cannot control the illumination of an exoplanet in this way. But the unavoidable rotation and revolution of any exoplanet will in general vary its phase, instellation, and the parts of its surface under illumination, and these effects can go partway toward this 'blur-free' strategy. Any mission to image an exoplanet at the SGL would need to account for these effects in its measurement strategy, and use or assume knowledge of the exoplanet's rotation speed and angle of inclination, and of its orbital inclination, eccentricity, argument of periapsis, and true anomaly. Finding an optimal strategy for exoplanet imaging at the SGL, and determining the required measurement time, while accounting for all these variables is well beyond the scope of this article. However, we can explore some restricted cases here.

An exoplanet orbit observed face-on will always present the exoplanet to the SGL at quarter phase. If this exoplanet were tidally locked to its host star, then the same region of the exoplanet would always be illuminated. For this specific case, the calculations presented earlier give a reasonable estimate for the time to image the illuminated region with a given resolution. But if the exoplanet were not tidally locked to its star, then the imaging strategy would need to account for the regions of the exoplanet going into and out of illumination. If the exoplanet rotation axis were also pointed directly at Earth, then the entire hemisphere facing Earth will be illuminated at some time during the orbit. To form a 1000x1000 pixel image of this hemisphere, we must still collect at least one million measurements (one for each pixel), but the amount of blur for each measurement will be reduced by about a factor of two (since half the exoplanet is always dark). This suggests a generalization of Equation \ref{eq:Ttotal}. Let $g$ be the fraction of the planet that is illuminated- $g=0.5$ for the case here. Then the blur contribution to a given pixel will be $gN$ times the directly imaged contribution, giving us the requirement that \( {\rm SNR}_R \ge {\rm SNR}_S \times gN\), and a total image measurement time of
\begin{equation}
T_{tot} \ge { {N^4 fQ_{cor} g^2 {\rm SNR}^2_S} \over {Q^2_{exo} }  }.
\end{equation}
We see therefore that for an exoplanet with these orbital parameters, being always at quarter phase ($g=0.5$) reduces the measurement time by a factor of four.

In general, the rotation axis of the exoplanet will not point directly towards Earth, and so more than one hemisphere is accessible for imaging. If an exoplanet could stay in full phase, while rotating with its axis pointing 90 degrees away from Earth, then if we choose to image both hemispheres with some resolution, it will take twice as long as imaging just the one hemisphere with that same resolution. We cannot hasten the imaging by choosing to observe only one hemisphere, since it will be facing away from Earth half the time anyway during these very long measurements.

We consider now the case of an exoplanet with both its orbital and rotation axes pointed 90 degrees from Earth, and not tidally locked. In this case, it will eventually present all of its surface to Earth, at all phases. We might extrapolate our result for the exoplanet at quarter phase to one at crescent phase. If only only 10\% of the exoplanet's surface is illuminated ($g=0.1$), then the total measurement time is reduced by a factor of 100, or 12 years for the 'exo-Earth at Proxima Centauri' case. This would double to 24 years if we chose to map the whole surface rather than just one hemisphere. One can imagine performing measurements with even thinner crescents to reduce the integration time even further, but there are a few difficulties with this approach. One is that an exoplanet can only have as little as 10\% of its surface illuminated as seen from Earth if its orbital plane is within 18 degrees of edge-on as viewed from Earth, which is relatively rare. Another is that the exoplanet would spend only a relatively small fraction of its orbit in such a crescent phase. Measurements during the rest of the orbit will be much less 'efficient' and do little  to reduce the total measurement time from what it would take to perform measurements only during the crescent phases. The total mission duration would be significantly lengthened by these 'waiting periods'. Finally, arbitrarily thin crescent phases are not accessible to this measurement. Exoplanets at close enough to new phase will actually be transiting their host stars. Near but outside transit, the blur from the SGL image of the much brighter host star will overlap with the light from the exoplanet itself, further corrupting the measurement. Both the transit and blur constraints will depend in general upon the exoplanet orbit and the size and luminosity of the host star.

\section{Conclusions}

The analysis presented in this article indicates that high-resolution imaging of an exoplanet by a meter-class telescope at the solar gravity lens will require extremely long integration times, due to the shot noise limited precision needed to separate the exoplanet signal from the coronal foreground and deconvolve the blurriness intrinsic to the SGL. Many simplifying assumptions were made to reach this conclusion. Better assumptions would lead to better estimates of the integration time. However, this analysis is not intended to give a precise estimate of the integration times for this type of mission, but rather to show that basic photometric considerations imply that a single telescope is very unlikely to produce a megapixel image within a human lifespan. Improving the estimate of integration time at the 50\% level would not significantly change this conclusion. Clearly, if a constellation of such telescopes were used rather than just one, with each telescope simultaneously measuring a different image pixel than the others, the total measurement time would potentially be reduced by a factor equal to the number of telescopes.

Furthermore, there are many potential factors not considered here that, if included in the analysis, could lead to even longer measurement times. For example, the visage that the exoplanet presents will not in general vary solely due to its illumination and rotation. The exoplanet may have clouds that form, move across its surface, and evaporate, and it may have seasons that make its albedo change with time due to freezing and thawing of water ice. If the image measurement time is long compared to the timescales for these processes, then the deconvolution method used here will lose accuracy.

Likely more troublesome is the impact if the solar corona foreground cannot be subtracted with photometric efficiency. For our case of an ${\rm SNR} = 10$ megapixel image of a full-phase exo-Earth at the same distance as Proxima Centauri, this photometric precision requires foreground corrected with 30 part-per-billion accuracy. Noise sources other than photon noise exceeding this level would dramatically increase required measurement times. 

\section{Acknowledgements}
The research reported here was carried out at the Jet Propulsion Laboratory, California Institute of Technology, under a contract with the National Aeronautics and Space Administration. The author is grateful to Mike Shao and Slava Turyshev of the Jet Propulsion Laboratory for helpful discussions. \copyright 2018 California Institute of Technology. Government sponsorship acknowledged.

\bibliographystyle{elsarticle-num}
\bibliography{SGLfinal}

\end{document}